\documentclass[english,aps,manuscript,reprint, twocolumn,superscriptaddress]{revtex4-1}
\usepackage[T1]{fontenc}
\usepackage[utf8]{inputenc}
\usepackage{graphicx}
\usepackage{dcolumn}
\usepackage{bm}
\usepackage{babel}
\usepackage{amsmath}
\usepackage{textcomp}
\begin{document}

\preprint{}

\preprint{Draft \#1}

\title{Differential Dynamic Microscopy microrheology of soft materials: a tracking-free determination of the frequency-dependent loss and storage moduli}

\author{Paolo Edera}

\affiliation{Dipartimento di Biotecnologie Mediche e Medicina Traslazionale, Università
degli Studi di Milano, Via. F.lli Cervi 93, Segrate (MI) I-20090,
Italy}

\author{Davide Bergamini}

\affiliation{Dipartimento di Biotecnologie Mediche e Medicina Traslazionale, Università
degli Studi di Milano, Via. F.lli Cervi 93, Segrate (MI) I-20090,
Italy}

\author{Veronique Trappe}

\affiliation{Department of Physics, University of Fribourg, Chemin du Musée 3,
CH-1700, Fribourg, Switzerland }

\author{Fabio Giavazzi}
\email{fabio.giavazzi@unimi.it}

\affiliation{Dipartimento di Biotecnologie Mediche e Medicina Traslazionale, Università
degli Studi di Milano, Via. F.lli Cervi 93, Segrate (MI) I-20090,
Italy}

\author{Roberto Cerbino}
\email{roberto.cerbino@unimi.it}

\affiliation{Dipartimento di Biotecnologie Mediche e Medicina Traslazionale, Università
degli Studi di Milano, Via. F.lli Cervi 93, Segrate (MI) I-20090,
Italy}

\date{\today}
\begin{abstract}
Particle tracking microrheology (PT-$\mu$r) exploits the thermal motion of embedded particles to probe the local 
mechanical properties of soft materials. Despite its appealing conceptual simplicity, PT-$\mu$r requires calibration procedures and operating assumptions that constitute a practical barrier to a wider adoption. Here we demonstrate Differential Dynamic Microscopy microrheology (DDM-$\mu$r), a tracking-free approach based on the multi-scale, temporal correlation study of the image intensity fluctuations that are observed in microscopy experiments as a consequence of the motion of the tracers. We show that the mechanical moduli of an arbitrary sample are determined correctly in a wide frequency range, provided that the standard DDM analysis is reinforced with a novel, iterative, self-consistent procedure that fully exploits the multi-scale information made available by DDM. Our approach to DDM-$\mu$r does not require any prior calibration, is in agreement with both traditional rheology and Diffusing Wave Spectroscopy microrheology, and works in conditions where PT-$\mu$r fails, providing thus an operationally simple, calibration-free probe of soft materials.
\end{abstract}
\maketitle

\section{Introduction\label{sec:Intro}}

Rheology is a well established experimental technique that probes the response of materials upon application of a stress or strain \citep{coussot2014rheophysics}. This probe is particularly significant for soft materials such as paint, starch, mayonnaise and gelatin that defy the sharp rules according to which we tend to classify a substance as an ideal viscous Newtonian liquid or a perfectly elastic Hookean solid. In practice, depending on the time scales probed and the magnitude of the stress or strain applied, soft materials may behave as solids or liquids. For instance, a dense colloidal suspension may exhibit a solid-like response upon application of small-amplitude, fast deformations whereas it may be more similar to a liquid upon applying  large-amplitude, slow deformations. Converting these important but qualitative considerations into some quantitatively and reproducibly determined mechanical moduli of the materials is the realm of rheology \citep{Wyss:2016eu}.

Traditional rheology makes use of rheometers, in which a soft material is loaded in the gap between two solid surfaces and stressed (or strained) in a controlled fashion to measure the strain (or stress) response of the material. This response can be entirely \footnote{As long as the stress or strain is small enough not to change materials' mechanical properties.} described in terms of a complex modulus $G^{*}(\omega)=G'(\omega)+iG''(\omega)$. $G^{*}$ can be measured with a rheometer, by imposing for instance an oscillatory strain $\gamma(t)=\gamma_{0}\sin(\omega t)$ and measuring the stress $\sigma(t)$ developed by the material. In general, one finds that $\sigma(t)=G'\gamma_{0}\sin(t)+G''\gamma_{0}\cos(t)$, where $G'$ and $G''$ are the storage (or elastic) and loss (or viscous) moduli of the material, respectively.

This denomination denotes that a Hookean solid is characterized only by a stress in
phase with the applied strain, with $G'$ corresponding to the elastic modulus of the solid, whereas the response of a Newtonian liquid is in quadrature, with $G''=\eta\omega$,
where $\eta$ is the dynamic viscosity. A generic soft material will have an in-phase response that can be associated to its solid-like character and an in-quadrature response that is due to its liquid-like nature.
Naively one can say that if for a given frequency $G'\gg G''$
the material is substantially a solid, whereas if $G'\ll G''$ it
behaves as a liquid. Inspecting the full frequency dependence of $G'$
and $G''$ provides thus a fundamental tool to classify materials based on their mechanical response or to monitor changes in their mechanical properties during for instance gelation or aggregation processes \citep{Kavanagh:1998fp}. 

Despite their powerfulness and immediacy, rheology tests performed with a
rheometer are affected by some limitations: they require a large quantity
of material (of the order of a few milliliters), they average over
possible heterogeneities of the sample, and the accessible frequency range is
limited at small $\omega$ by torque limitations and at large
$\omega$ by inertial effects \citep{Ewoldt:2015qr}.

A complementary approach that addresses the above issues, is represented
by microrheology \citep{Cicuta:2007vn,Mason:1997yu,Squires:2010hb,Waigh:2005jk,Waigh:2016sf}.
Originally introduced by Mason and Weitz in 1995 \citep{Mason:1995fq},
the so-called passive microrheology consists of seeding the soft material
of interest with tracer particles of radius $a$ and measuring the
mean-square displacement (MSD) $\left\langle \Delta r^{2}(t)\right\rangle $
of the tracers within the material as a function of time $t$. The MSD can be related to the frequency-dependent complex
modulus $G^{*}(\omega)$ by using the generalized
Stokes-Einstein equation \citep{Squires:2010hb}
\begin{equation}
G^{*}(\omega)=\left.\frac{dk_{B}T}{3\pi as\left\langle \Delta\tilde{r}^{2}(s)\right\rangle }\right|_{s=i\omega}\label{eq:GSE}
\end{equation}
where $d$ is the number of dimensions tracked in the MSD, $k_{B}$
the Boltzmann constant, $T$ the temperature, $i$ the imaginary
unit and $\left\langle \Delta\tilde{r}^{2}(s)\right\rangle $ the
Laplace transform of the MSD. In a
Newtonian liquid of viscosity $\eta$, the MSD
of a tracer particle with diffusion coefficient $D_0=k_{B}T/(6\pi\eta a)$
is given by $\left\langle \Delta r^{2}(t)\right\rangle =2dD_0t$, which
leads to $\left\langle \Delta\tilde{r}^{2}(s)\right\rangle =2dD_0/s^{2}$
and, in turn, to $G^{*}(\omega)=iG''(\omega)=i\omega\eta$. For a
solid, instead, the elastic modulus $G^{*}(\omega)=G'$ is obtained
from estimating the mean squared displacement $\left\langle \Delta r^{2}(t)\right\rangle =\frac{dk_{B}T}{3\pi aG'}$
of a particle in an elastic trap with the condition $\left\langle \Delta r^{2}(t)\right\rangle =0$
for $t<0$. The Laplace transform is then given by $\left\langle \Delta\tilde{r}^{2}(s)\right\rangle =\frac{dk_{B}T}{3\pi aG'}\frac{1}{s}$. 

The MSD of the tracer particles can be obtained in a variety of ways \cite{Waigh:2005jk,Cicuta:2007vn,Squires:2010hb}. A direct way is to track in real-space the trajectories of the tracer particles, as done in Particle Tracking (PT) experiments \citep{Mason:1997yu}. An alternative way is to extract the MSD from the measurement of the intensity scattered or fluorescently emitted
by a dilute collection of non-interacting tracer particles, as done
in Diffusing Wave Spectroscopy (DWS) \citep{Mason:1995fq}, Dynamic
Light Scattering (DLS) \citep{mason1996rheology,Dasgupta:2002jt} and Fluorescence Correlation
Spectroscopy (FCS) \citep{Rathgeber:2009mb}. Historically, DWS-$\mu$r
was the first to be developed and, together with PT-$\mu$r, is still one
of the most common approaches. DWS and PT
are in principle quite complementary. DWS gives access to short time
scales and small MSD, while PT gives access to longer time scales and larger MSD. However, while DWS can be used with almost no user intervention, PT involves a rather tedious and delicate selection of the
trajectories. The disadvantage of DWS in turn is to require larger tracer particle concentrations that may more easily alter the mechanical properties of the material itself.

Almost ten years ago, the usefulness of a technique named Differential Dynamic Microscopy (DDM) was demonstrated for the characterization of the dynamics of colloidal suspensions of particles \citep{Cerbino:2008if}.
One of the main features of DDM is that it lies somehow in-between
PT and DLS. Similar to PT, it is based on real-space movies collected
in microscopy experiments. These data are treated via an image processing
algorithm \citep{Croccolo:2006et} or equivalent versions of it \citep{Giavazzi:2014sj}
that combines image differences and spatial Fourier transformations
to obtain as a result the intermediate scattering function $f(\mathbf{q},t)$
that is typically probed in DLS experiments as a function of the scattering
wave-vector $\mathbf{q}$ and time $t$ \citep{Giavazzi:2009xd}.
Since its introduction, DDM has been profitably used and extended
also by several groups \citep{Wilson:2011wa,Reufer:2012ri,He:2012bx,Gao:2015zh,Germain:2015gf,Dehaoui:2015mz,Wittmeier:2015qy,Sentjabrskaja:2016fq,Wulstein:2016tt,Kodger:2017mw}
for a variety of applications. Surprisingly, as of today, there have been no attempts to 
use DDM to perform microrheology experiments
on complex fluids, which would seem a quite obvious and powerful application.

In this work, we show that DDM can indeed be used as a convenient and reliable tool to probe the mechanical properties of complex fluids, which we demonstrate with both Newtonian liquids, obtained by mixing water and glycerol in variable proportions, and viscoelastic samples, consisting of aqueous solutions of a high molecular weight polymer (polyethylene oxide). To determine the tracer MSD in DDM, we demonstrate the advantage of a novel fitting-free, optimization-based procedure that is applicable to an arbitrary sample and does not require any prior calibration. The obtained results are found to be in agreement with standard rheology and with both PT- and DWS-microrheology. In addition, we show that DDM-$\mu$r operates also with small tracer particles that are not suitable for tracking experiments; this widens the range of applicability of microrheology.

Our results show that optimization-based DDM-$\mu$r is a flexible, calibration-free approach to microrheology that can be almost fully automated, thus eliminating the arbitrariness, typical of PT experiments, in sorting and selecting the suitable trajectories. We expect that DDM-$\mu$r can be successfully used to measure the rheological properties of a variety of soft materials, also in cases where DLS, DWS and PT can not be used. A typical example is the cell interior \cite{drechsler2017active} where DDM has already been successfully used to measure the interplay of diffusion and flow during oogenesis.

\section{Materials and Methods}\label{mandms}

\subsection{Samples preparation\label{sub:Sample-preparation}}
We used two different classes of samples: Newtonian
fluids with varying viscosity obtained by adding different amounts of glycerol to water, 
and viscoelastic fluids consisting of aqueous solutions of polyethylene oxide (PEO, $M_{W}=2\times10^{6}$ Da). For the PEO solution we chose to work at a concentration of c=2.0 wt\% above the overlap concentration (c{*}=0.09 wt\%) to obtain a sample with appreciable
viscoelastic properties in the frequency range of interest. At these conditions, the mesh size of the polymer network is estimated to be $\sim 10$ nm \cite{Dasgupta:2002jt}.

\subsubsection*{Glycerol-water solutions}
The glycerol-water samples were prepared by mixing suitable amounts
of glycerol (Sigma Aldrich), MilliQ water and an aqueous suspension of latex beads (Sigma Aldrich, LB5, nominal diameter 0.45-0.47 $\mu m$, solid content 10\%) to reach final mass fractions of glycerol equal
to $0$\%, $48.8$\%, $82.7$\% and $97.5$\%. The mass fractions
of suspended beads were $0.075$\%, $0.14$\%, $0.18$\% and $0.2$\%,
respectively. All four samples were investigated by using both PT and DDM.

\subsubsection*{Polymer solution}
The PEO solution for traditional rheology experiments was prepared from the pure product purchased as powder (Sigma Aldrich, prod. code 372803). The powder was carefully dissolved in MilliQ water previously filtered with membrane filters (pore size 0.2 $\mu m$). To prevent the formation of clumps of undissolved polymer, water was gradually added to the polymer while stirring. The solution was then kept in incubation for nine days at about 40 \textdegree{} C. A few drops of a Sodium Azide solution (molarity 4 mM) were added to the PEO solution to prevent bacterial proliferation.

For the DDM, PT and DWS experiments, in which colloidal tracer particles were to be added to the PEO solution, a prescribed amount of pure water was replaced with the aqueous colloidal suspensions of latex beads during sample preparation. The beads were purchased by Sigma Aldrich with the part numbers LB1 (nominal diameter 0.10-0.12 $\mu m$, solid content 10\%) and LB5 (nominal diameter 0.45-0.47 $\mu m$, solid content 10\%). The bead sizes were also tested with DDM and were found to be equal to $0.112\pm0.002$ $\mu m$ and $0.445\pm0.005$ $\mu m$ for the LB1 and LB5 samples, respectively. After dilution the final concentration of PEO was 2\% wt/wt and
the final concentration of tracer beads was $(1.00\pm 0.05)\times 10^{-3}$ wt/wt. To ensure multiple scattering for the DWS experiment we used LB5 particles at $(1.00\pm 0.05)\times 10^{-2}$ wt/wt.

\subsection{Rheology\label{sub:Rheology}}
Since for the Newtonian glycerol-water mixtures reliable literature data are available, standard rheology experiments were only performed for the PEO sample. We used a commercial rheometer (Anton Paar MCR502) equipped with cone and plate geometry (radius=$25$ mm, cone angle = $1^{o}$) to apply an oscillatory shear strain with strain amplitude of 5\%, and angular frequency in the range $[0.1,100]$ s$^{-1}$. Our experiments were perforemed in the temperature range $T=20-25$ \textdegree C. To avoid evaporation during measurement we used a solvent trap.

\subsection{Particle tracking (PT)\label{sub:Particle-tracking}}
PT experiments were performed by tracking LB5 particles dispersed in the four glycerol-water samples and in a polymer solution with the PEO concentration also used in \ref{sub:Rheology}.
The samples were loaded in a capillary (Vitrocom) with rectangular cross
section and internal dimensions $10\times2\times0.1$ mm$^{3}$. Microscopy experiments were performed  in bright field with an optical microscope (Nikon Ti-E), equipped with a digital camera (Hamamatsu
Orca Flash 4.0 v2), and a 40x objective. The resulting pixel size was $d_{pix}=162.5$ nm. Image sequences made of $10,000$ images (512x256 pixels) were acquired
at two different frame rates (777 Hz and 10 Hz). In all acquired images, the sample appeared transparent and the colloidal particles were clearly visible.

Particle tracking analysis was conducted by using a customized version of the MATLAB code script made freely available by the group of Maria Kilfoil at UMass (people.umass.edu/kilfoil/). This software reconstructs the individual trajectories of several particles in parallel, calculating their MSD as a function of time. Compared to the original code, we added some custom features, mainly to adapt our analysis to bright-field time-lapse movies and to estimate error bars and experimental uncertainties.

Once the MSD was obtained, the data needed to be corrected by subtracting the additive contribution due to the intrinsic localization uncertainty that becomes dominant for small times and particle displacements \cite{crocker1996methods}. This step, which lies at the core of PT-$\mu$r, requires an independent calibration of the particle localization error. In our experiments, the static localization error was determined as that which minimizes the deviation from a purely linear behavior in the MSD measured in Newtonian samples \cite{Waigh:2005jk}.

Once the corrected MSD was obtained, we followed different procedures for the two classes of samples. Results for MSD of the Newtonian fluids were simply fitted to a straight line and the sample viscosity $\eta$ was obtained from the slope,  $4k_{B}T/(6\pi\eta a)$. For the PEO sample, we used the Kilfoil-group software to extract the frequency-dependent elastic and loss moduli, $G'$ and $G''$, respectively.
The software implements an algebraical inversion procedure based on the work of T.G. Mason \textit{et al.} \cite{Mason:1997yu}.


\subsection{Diffusing Wave Spectroscopy (DWS)\label{sub:Diffusing-Wave-Spectroscopy}}

DWS microrheology experiments were performed on a polymer solution with the PEO concentration also used in \ref{sub:Rheology}. The tracer particles concentration (1\%) was chosen to ensure multiple scattering \cite{Mason:1995fq}. In the limit of multiple scattering the autocorrelation function
of the scattering intensity is given by
\begin{equation}
g(\tau)=\int^{\infty}_{0} P(s)e^{-\frac{k^{2}_{0}}{3}\left<\Delta r^{2}(\tau)\right>\frac{s}{l^{*}}}ds
\label{eq:dws}
\end{equation}
where $k_{0}=2\pi n /\lambda_{0}$ is the wave-vector of light with wavelength $\lambda_{0}$ (in our experiment $687$ nm) incident on a medium with refractive index $n$. $P(s)$ is the scattering-geometry-dependent relative probability distribution of photon path lengths $s$ inside the medium and $l^{*}$ is the transport mean free path, which quantifies the distance that a photon has to travel inside the sample before loosing memory of its original direction. For our sample, we found $l^{*}=256\pm5$ $\mu$m. The MSD can be extracted by inverting Eq. \ref{eq:dws} \cite{pine_book}. 

For our experiments we used the commercial instrument
\textit{DWS Rheolab} (LSInstruments, Fribourg, Switzerland), a compact stand-alone optical microrheometer that is based on DWS. The sample was hosted in a cuvette of thickness $L=2$ mm. Measurements were performed in transmission geometry with a duration of 3000 s each. The MSD of the tracer particles was obtained in the time range $1.38$ $\mu$s$-0.71$ s and subsequently analyzed to extract the moduli $G'$ and $G''$.
%
%

\begin{figure*}
\centering
\includegraphics[width=.99\textwidth]{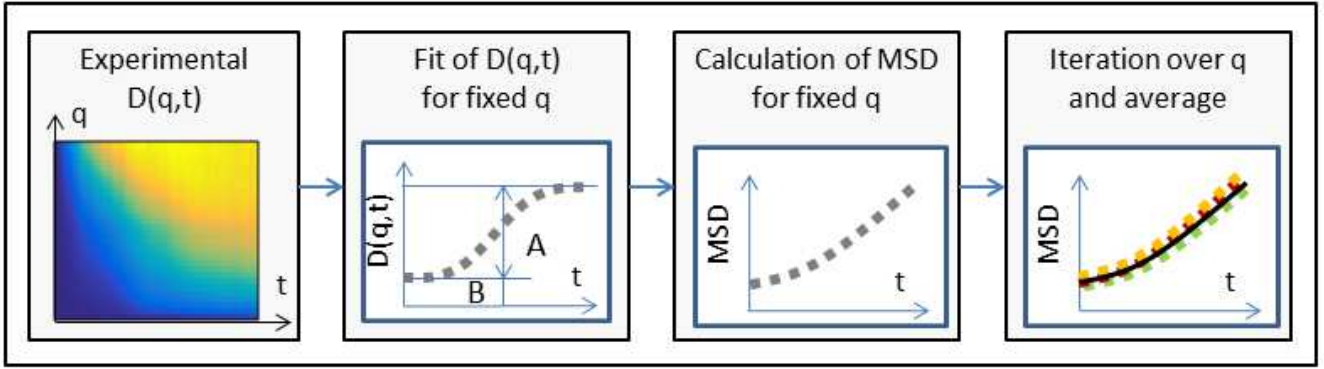}
\caption{Schematic representation of the fitting-based procedure used to determine MSD from the image structure function. For a given wave-vector $q$, the image structure function $D(q,t)$ is fitted to a model to obtain the noise baseline $B(q)$ and the signal amplitude $A(q)$. Using Eq. \ref{eq:msdddm2}, an estimate for the MSD is obtained. The procedure is then repeated for all $q$-values in the selected $q$-range and the best estimate for the MSD is obtained as the average of the curves obtained for all $q$-values.}
\label{fig:diagram0}      
\end{figure*}

\subsection{Differential Dynamic Microscopy (DDM)\label{sub:Differential-Dynamic-Microscopy}}

We performed DDM measurements on all samples. LB5 particles were used for the four Newtonian samples and for the PEO solution. The latter was also studied with LB1 particles; for this sample  PT is not feasible. Standard DDM analysis was based on a repeated sequence of image subtractions
and image Fourier transforms
\cite{Cerbino:2008if,Giavazzi:2009xd,Giavazzi:2014sj}. In more detail, the
image structure function for all the accessible two-dimensional wave-vectors
$\mathbf{q}=\left(q_{x},q_{y}\right)$ was calculated for a set of
time delays $t$ according to
\begin{equation}
D\left(\mathbf{q},t\right)=\left\langle \left|I\left(\mathbf{q},t_{0}+t\right)-I\left(\mathbf{q},t_{0}\right)\right|^{2}\right\rangle _{t_{0}}\label{eq:strfun}
\end{equation}
where $I\left(\mathbf{q},t\right)$ is the Fourier transform of the
image $I\left(\mathbf{x},t\right)$ acquired at time $t$ in a fixed
plane in which the horizontal position is labeled by $\mathbf{x}=\left(x,y\right)$. It has been recently shown that multiplying the images with a windowing function before performing the Fourier transform operation removes the artifacts due to the finite image size and improves the determination of $D\left(\mathbf{q},t\right)$, especially for those $q$ for which the signal is comparable or smaller than the noise \cite{2017arXiv_apodization}. We thus apply this algorithm in our analysis. 

The image structure function is quantitatively related to the normalized
intermediate scattering function $f\left(\mathbf{q},t\right)$ and
in most cases of interest the simple relationship
\begin{equation}
D\left(\mathbf{q},t\right)=A(\mathbf{q})\left[1-f\left(\mathbf{q},t\right)\right]+B(\mathbf{q})\label{eq:strfun2}
\end{equation}
holds, where the functions $A(\mathbf{q})$ and \textbf{$B(\mathbf{q})$},
usually treated as fitting parameters, are set by the spatial intensity
correlations and the noise of the detection chain, respectively. The
normalized intermediate scattering function has some general properties
such that $f\left(\mathbf{q},0\right)=1$ and $f\left(\mathbf{q}, t\rightarrow\infty\right)=0$
if the particles position are fully uncorrelated for long times \cite{Berne:2000ye}. For a dilute collection of non-interacting particles one has
\begin{equation}
f\left(\mathbf{q},t\right)=e^{-\frac{q^{2}}{4}\langle\Delta r^{2}(t)\rangle}\label{eq:msdddm}
\end{equation}
which is the two-dimensional equivalent of the main assumption on
which DLS microrheology is based \cite{Mason:1997yu,Dasgupta:2002jt}. Thus, under conditions in which Eq. \ref{eq:msdddm} holds,
the MSD of tracer particles dispersed in a soft material can be obtained
as

\begin{equation}
\langle \Delta r^{2}(t) \rangle=-\frac{4}{q^{2}}\ln\left(1-\frac{D\left(\mathbf{q},t\right)-B\left(\mathbf{q}\right)}{A\left(\mathbf{q}\right)}\right).\label{eq:msdddm2}
\end{equation}

provided that an accurate fitting of the structure functions can be performed, as sketched in Fig. \ref{fig:diagram0}. Once the MSD is obtained it can be used to estimate the loss and elastic moduli of the sample, which is done here with the same tools used to treat PT data.

We note that even though the whole procedure to extract the MSD from the DDM analysis of microscope movies appears at first rather straightforward, a successful and accurate output requires
the precise knowledge of $A\left(\mathbf{q}\right)$ and $B\left(\mathbf{q}\right)$.
When an accurate fitting model for $\langle \Delta r^{2}(t) \rangle$ is available, as this is the case for freely diffusing particles where $\langle \Delta r^{2}(t) \rangle=4D_{0}t$, this can be done also if the key experimental parameters
(image exposure time, acquisition frame rate, total number of images)
do not allow to observe the full relaxation of the intermediate
scattering function from one to zero. By contrast, if such a model is not available any spurious effect altering the determination of
$A$ and $B$ will impair the determination of a correct MSD. Clearly,
if DDM is to be used as a general purpose probe of the mechanical properties of soft materials, suitable precautions
need to be taken to guarantee a model-free determination of the MSD.
In Section \ref{model-inde}, we will show how this task can be accomplished
by replacing the standard fitting-based DDM analysis with a suitable optimization-based DDM analysis.

\section{Results and discussion}

\subsection{Newtonian fluids: fitting-based DDM analysis}
The DDM experiments presented here are aimed at measuring the viscosity of the Newtonian fluids seeded with $\simeq 0.5$ $\mu m$ latex beads (LB5) described in detail in Section \ref{mandms}. The key step of the analysis consists in extracting the MSD of the tracers directly from the image structure function by using Eq. \ref{eq:msdddm2}. To this aim the amplitude $A(\mathbf{q})$ and the baseline $B(\mathbf{q})$ need to be known with high accuracy for each $q$, as any systematic error in their determination would introduce a bias in the MSD. In particular, an overestimate (underestimate) of $B(\mathbf{q})$ would lead to a spurious acceleration (deceleration) of the reconstructed tracers dynamics for small times.

For monodisperse non-interacting colloidal particles dispersed in a Newtonian fluid, the intermediate scattering function decays exponentially $f\left(\mathbf{q},t\right)=e^{-D_0q^{2}t}$ and $A(\mathbf{q})$ and $B(\mathbf{q})$ can be simply obtained by fitting the image structure functions to Eq. \ref{eq:strfun2}.

\begin{figure}
\centering
\includegraphics[width=.45\textwidth]{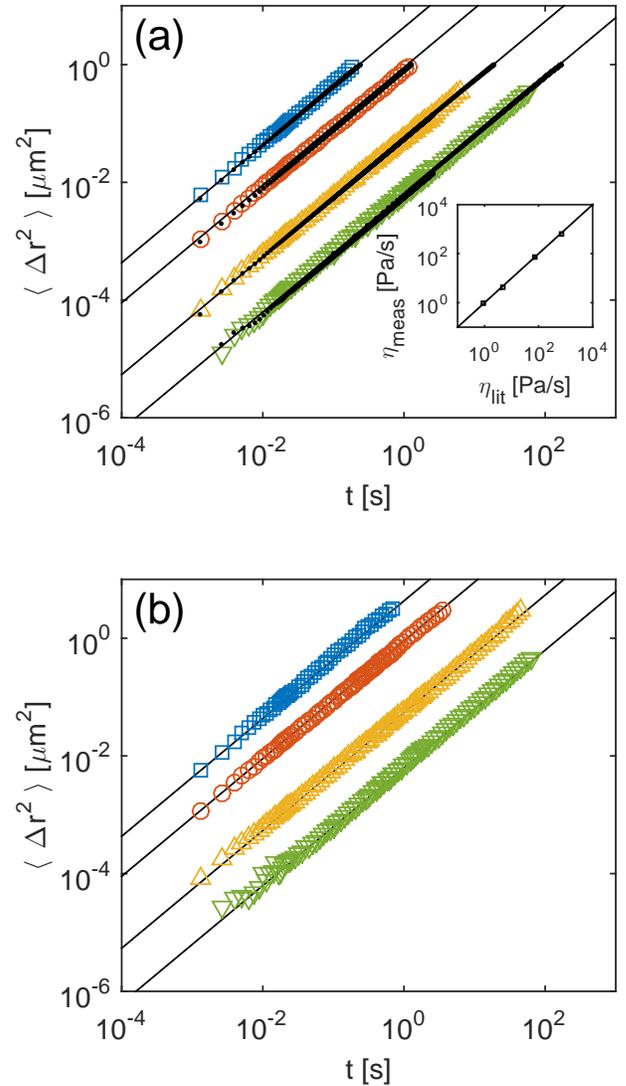}
\caption{MSD of LB5 tracers in different water-glycerol solutions. (a) symbols: MSD obtained from DDM with the fitting-based procedure (details in main text); dots: MSD obtained from PT; continuous lines: expected MSD from Ref. \cite{Glycerolwater}. The viscosities $\eta_{meas}$ experimentally determined with DDM are shown in the inset as a function of the expected values $\eta_{lit}$ (symbols). The error bars are smaller than the symbols and the continuous line corresponds to the identity $\eta_{meas}=\eta_{lit}$. (b) symbols: MSD obtained from DDM with the model-free procedure (see main text for details), continuous lines: expected MSD, as in panel a).}
\label{fig:newton}      
\end{figure}

In our experiments, we found indeed that for all samples the intermediate scattering functions were very well described in terms of a single exponential relaxation for all the wave-vectors in the range of $[2.27, 9.82]$ $\mu m^{-1}$ for pure water, of $[2.27, 9.06]$ $\mu m^{-1}$ for $48.8$\% glycerol in water, of $[3.02, 9.06]$ $\mu m^{-1}$ for $82.7$\% glycerol in water, and of $[3.78, 9.06]$ $\mu m^{-1}$ for $97.5$\% glycerol in water. In practice, the width of the wave-vector range is set by the $q$-region in which both $A(q)$ and $B(q)$ are known accurately and Eq. \ref{eq:msdddm2} can be used to obtain the MSD from the intermediate scattering functions.
For each wave-vector in the range $[3.78, 9.06]$ $\mu m^{-1}$ we thus extracted an estimate for the MSD. These estimates were then combined to obtain a $q$-averaged estimate of the MSD for all the samples. These MSD are reported in Fig.\ref{fig:newton}a for the four Newtonian samples investigated here. All the curves are in excellent agreement with the PT results obtained by analyzing the same image sequences. For the viscosity we obtain $\eta_{meas}=0.92 \pm 0.03,\,4.4 \pm 0.2,\, 73 \pm 3,\, 633 \pm 40 $ mPa s. These values are in very good agreement with those expected $\eta_{lit}=0.914 \pm 0.01,\, 4.9 \pm 0.05,\, 69.2 \pm 0.7,\, 702 \pm 50 $ mPa s, \cite{Glycerolwater}, as shown in the inset of Fig. \ref{fig:newton}(a). The experimental uncertainty on the value obtained for $97.5$\% glycerol in water is due to the uncertainty in the sample composition.

\subsection{Newtonian fluids: optimization-based DDM analysis}\label{model-inde}
As we will show in the following, the satisfactory results obtained by using the standard DDM analysis with Newtonian samples depended on the fact that a model for the intermediate scattering function was readily available. In general, this would not be the case, as the behavior of a generic soft material is not known \textit{a priori}. For this reason, we devised a simple, self-consistent procedure that exploits the multi-$q$ capability of DDM to extract the MSD of tracer particles for an arbitrary sample in a robust way. The proposed procedure builds on the automatic determination of $A(q)$ and $B(q)$ based on an iterated optimization cycle.

\begin{figure*}[!th]
\centering
\includegraphics[width=.99\textwidth]{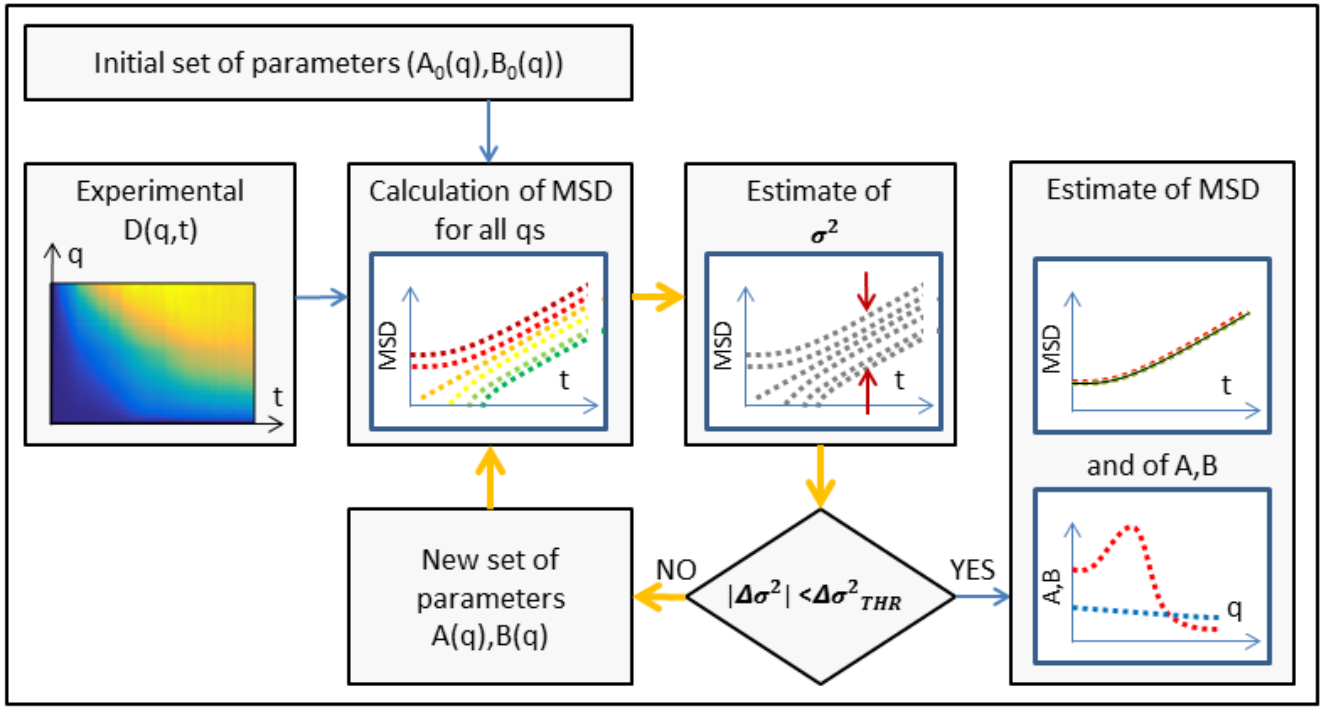}
\caption{Schematic representation of the optimization-based DDM analysis. The procedure is based on an optimization cycle (yellow arrows), fed by the experimental image structure  function $D(q,t)$ and by an initial set of parameters $\left( A_0(q), B_0(q) \right)$. The object function is the dispersion $\sigma^2$ of the reconstructed mean square displacements (see Eq. \ref{appsigma2}). New values of $\left( A(q), B(q) \right)$ are iteratively generated in order to minimize the object function. The output of the procedure is the optimal set of parameters $\left( A(q), B(q) \right)$ leading to the best estimate of MSD($t$).} 
\label{fig:diagrams}      
\end{figure*}

The general idea is sketched in Fig.\ref{fig:diagrams}, where we show a block-diagram that depicts our fitting-free 
procedure. This procedure is based on an optimization cycle initially fed by a tentative amplitude-baseline parameter pair $\left(A_0(q),B_0(q)\right)$, for $q$-values within a given interval $[q_1,q_2]$. These parameters are used to invert the corresponding image structure functions (Eq. \ref{eq:msdddm2}), leading to a "bundle" of MSDs. If the considered pair $\left(A_0(q),B_0(q)\right)$ is the correct one for all $q$s, than the estimates for the MSD given by Eq. \ref{eq:msdddm2} are completely $q$-independent, resulting in an almost perfect collapse of all the curves. Any deviation of the parameters from the correct values introduces a $q$-dependent dispersion.
In our optimization scheme, the dispersion $\sigma^2$ of the curves (see Appendix A for details) plays the role of an objective function: new values of $\left(A(q),B(q)\right)$ are iteratively generated until a minimum of $\sigma^2$ is found.
This algorithm, implemented in a custom code developed in MATLAB\textregistered, was found to rapidly and robustly converge to a minimum for a wide range wave-vectors.

Results obtained for the tracer MSD with this optimization-based procedure (Fig. \ref{fig:newton}b) are in excellent agreement with those obtained with the fitting-based analysis (Fig. \ref{fig:newton}a) over the whole investigated range of delay times $1.3 \times 10^{-3}$ $s< t<10^2$ $s$, which validates the procedure. Also, we note that the $q$-averaged MSD shown in Fig. \ref{fig:newton}b were obtained by averaging the MSD in the range $1.36$ $\mu m^{-1}<q<9.06$ $\mu m^{-1}$; this range is wider than that probed with the fitting-based procedure. The usable $q$-range is larger in the optimization-based procedure, because  the full relaxation of the image structure functions is here not a requirement for the determination of the MSD, since $A(q)$ and $B(q)$ can be obtained self-consistently. Let us stress that the optimization-based procedure is largely model- and operator-independent. The only required external parameters are the relevant $q$-range $[q_1,q_2]$ over which the optimization is performed and the initial values of the parameters $(A_0(q),B_0(q))$.

The key importance of all these properties when studying arbitrary samples is described in detail in the next subsection.

\subsection{Viscoelastic fluid}
\begin{figure}[!t]
\centering
\includegraphics[width=.5\textwidth]{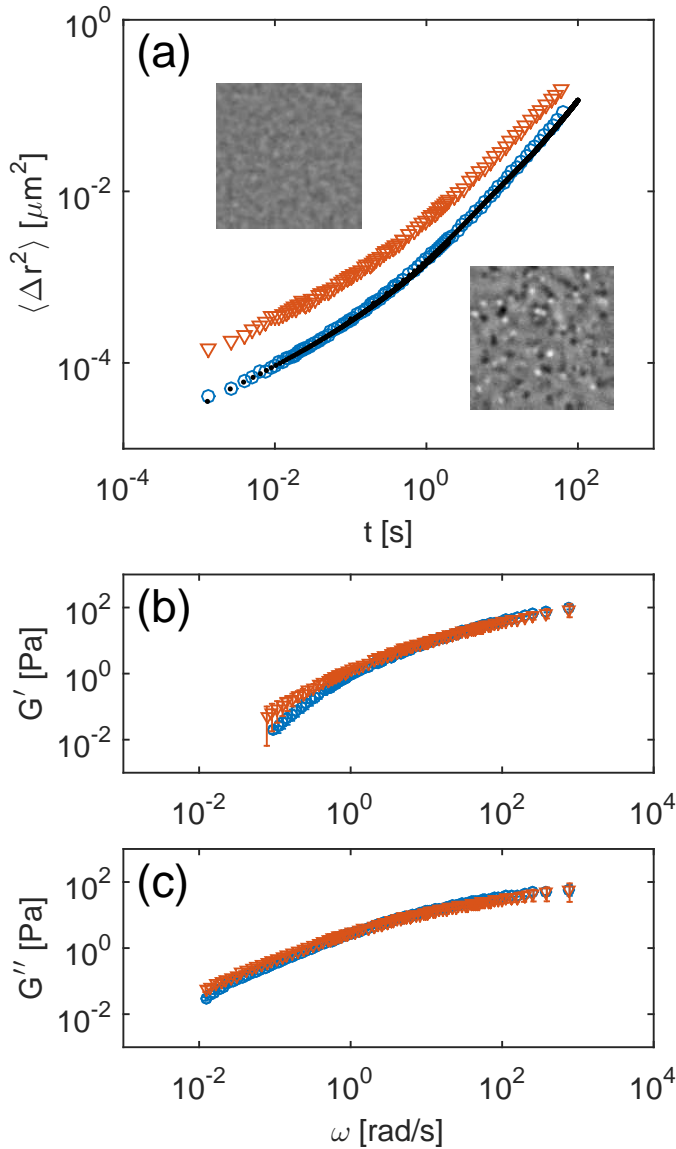}
\caption{(a) Two-dimensional MSD of LB1 (orange triangles) and LB5 (blue circles) tracers in a viscoelastic polymer solution (2\% PEO2 in water) obtained from DDM analysis. Black dots: same quantity obtained from PT analysis for the sample with LB5 tracers. The small insets show representative images of the two samples: the one loaded with subdiffraction LB1 particles (left upper corner) and the one loaded with LB5 tracers (right lower corner), respectively. (b) Comparison of the storage moduli $G^{\prime}$ estimated from the DDM-reconstructed MSD of LB1 tracers (orange triangles) and LB5 tracers (blue circles), respectively.
(c) same as in panel (b) for the loss moduli $G^{\prime \prime}$.}
\label{fig:PEO2_MSD}      
\end{figure}

\begin{figure}[!t]
\centering
\includegraphics[width=.5\textwidth]{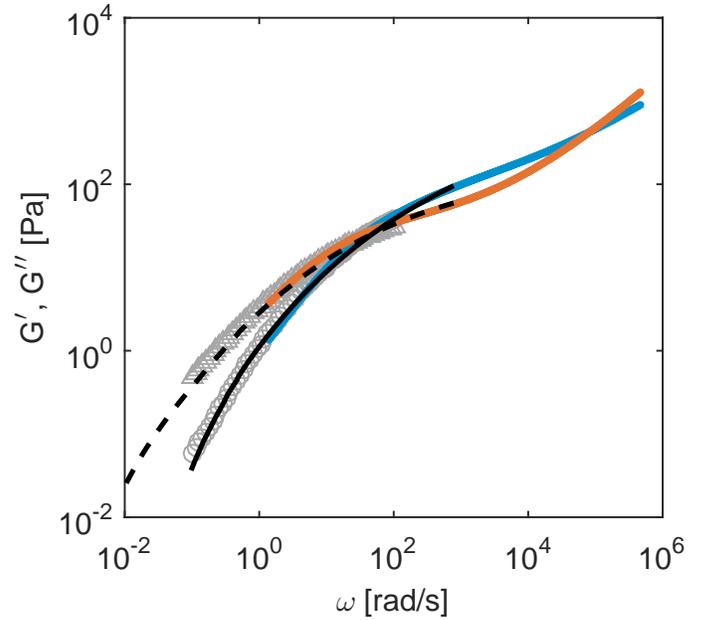}
\caption{Comparison of the viscoelastic moduli $G^{\prime}$ and $G^{\prime \prime}$ of a 2\% PEO polymer solution in water, obtained with different methods. Gray circles (triangles): $G^{\prime}$ ($G^{\prime \prime}$) obtained with traditional rheology; continuous blue (orange) line: $G^{\prime}$ ($G^{\prime \prime}$) obtained with DWS using LB5 tracers; black continuous (dashed) line: $G^{\prime}$ ($G^{\prime \prime}$)  obtained with DDM microrheology (weighted average of the results of LB1 and LB5 tracers, shown individually in  Fig. \ref{fig:PEO2_MSD}).}
\label{fig:PE2_rheo}      
\end{figure}

In this subsection, we apply the optimization-based procedure to the data obtained with our model viscoelastic fluid, an aqueous solution of PEO, that exhibits elastic behavior at short-times, high frequencies.
The expected short time elastic plateau in the MSD would contribute to the baseline $B(q)$, requiring an independent determination of the camera noise. Such requirement would be similarly involved as the calibration procedure needed in PT-$\mu$r experiments to account for the tracer localization uncertainty. While such calibration is technically feasible, the optimization-based DDM analysis permits a calibration-free implementation of DDM-$\mu$r.

Application of the optimization-based procedure to the PEO solutions with small ($\sim 100$ nm) and large ($\sim 500$ nm) tracers provides the results shown in Fig. \ref{fig:PEO2_MSD}a. The accessible range of probed timescales is very similar for the two tracer sizes. For comparison we also show the results obtained with PT for the sample containing the larger tracers as black points in Fig. \ref{fig:PEO2_MSD}a; for the smaller particles tracking is not feasible, as easily appreciated from the images shown as insets.

\begin{figure*}[!t]
\centering
\includegraphics[width=.99\textwidth ]{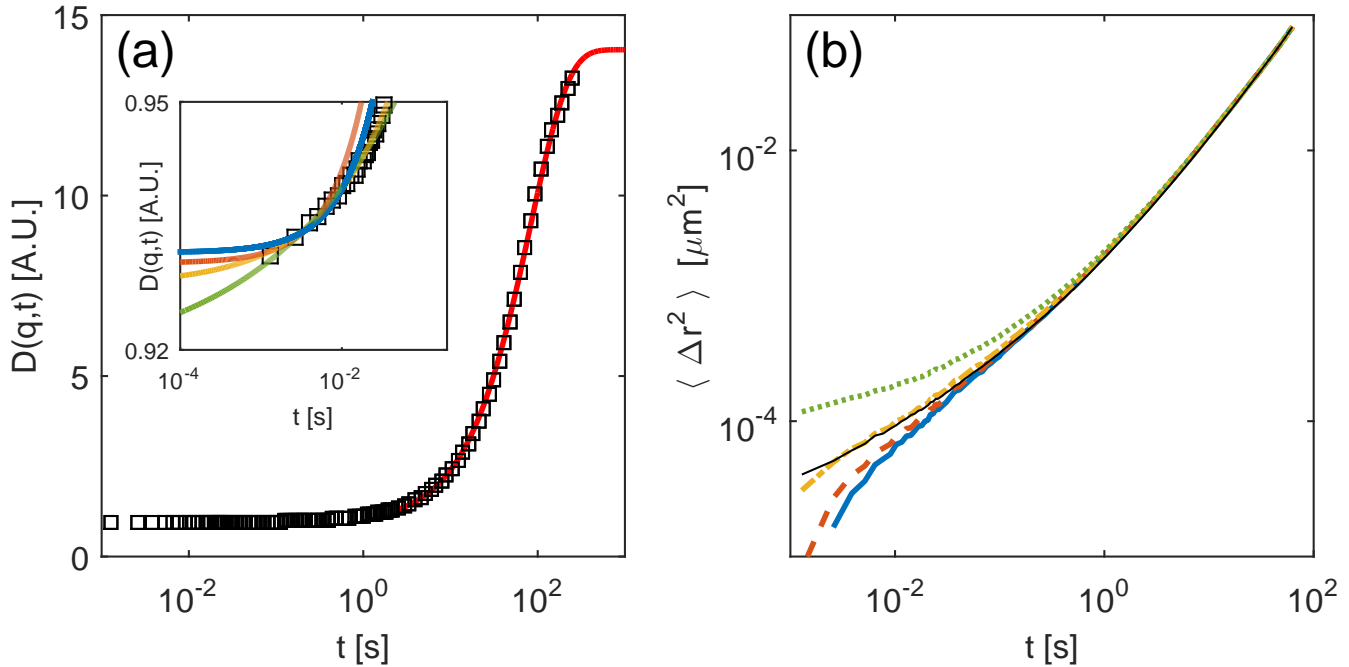}
\caption{Effect of the model-dependent determination of the noise baseline $ B(q)$ on the reconstructed MSD. (a) Symbols: image structure function $D(q,t)$ (for $q=6.04$ $\mu m^{-1}$) obtained from DDM on LB5 tracers in a viscoelastic solution of PEO in water. The continuous red line is an exponential fit to the data obtained at large delay times ($t>20$ s); this fit allows for the estimation of the plateau height $A^{\prime}(q)$. Inset: close-up of the short time behavior of $D(q,t)$ (symbols). The data is fitted with different functions, leading to different estimates of the baseline $B(q)$: linear fit over the first $10$ data points (continuous blue line), linear fit over the first $5$ data points (dashed orange line), fit over the first 20 data points with a function of the form $y=ax^{0.5} +b$ (dashed-dotted yellow line) and fit over the same interval with a function of the form $y=ax^{0.25} +b$ (dotted green line). (b) Mean square displacement obtained from Eq. \ref{eq:msdddm} using the amplitude $A(q)=A^{\prime}(q)-B(q)$ and the noise baseline $B(q)$ obtained from the different fitting models shown in the inset of panel (a). Curves are color-coded according to the fits, the black continuous line is the result of the model-free procedure shown in Fig. \ref{fig:PEO2_MSD}(a) as blue circles.}
\label{fig:PEO2rightandwrong}      
\end{figure*}
For each tracer size, we extracted from the MSD the mechanical moduli $G^{\prime}$ and $G^{\prime \prime}$, as shown in Fig. \ref{fig:PEO2_MSD}(b) and (c). Results obtained for $G^{\prime}$ with the two tracer sizes are off by about 10-20\% at small frequencies but the two datasets are compatible within the experimental errors. To obtain a statistically significant estimate for the moduli, we combine the data obtained with the two tracers and show the results as black lines in Fig. \ref{fig:PE2_rheo}. These data are in good agreement with the results obtained with traditional rheology, shown as open symbols, and also with the results obtained with DWS, shown as closed symbols. DDM-$\mu$r extends traditional rheology by one decade at high frequency, whereas at low frequency similar performances are obtained, at least as far as the storage modulus is concerned. However, improvements in the low-frequency region may be expected by increasing the mechanical stability of the microscope setup.

Let us underline that, without additional calibration steps, it would be very difficult to extract meaningful MSD and thus mechanical moduli with a fitting-based analysis of the DDM-data. In the limit of short times the MSD displays a non-trivial scaling, compatible with a power-law $MSD \simeq  t^\gamma$ with an exponent $\gamma$ close to $0.5$. The counterpart of this behavior in the Fourier space is an image structure function taking the form of $D(q,t) \simeq C(q) t^{\gamma}+B(q)$ at short times. Clearly, an exponential or a polynomial fit of $D(q, t)$ are inadequate to describe this behavior and any estimate of the baseline $B(q)$ based on an exponential or a polynomial fit provides a biased, incorrect result, as shown in Fig. \ref{fig:PEO2rightandwrong}. Choosing other model functions, such as for instance a power-law with different exponents, also fails, even though the data may seem deceivingly well described at short times. By contrast, the optimization-based procedure self-consistently determines the MSD without need of fitting the experimentally determined image structure functions.

\section{Conclusions}
Microrheology is a very powerful complement to traditional, mechanical rheology \cite{Waigh:2005jk,Cicuta:2007vn,Squires:2010hb,Waigh:2016sf}. For the high-frequency range, rheology is usefully complemented by DWS-$\mu$r \cite{Mason:1995fq}, whereas in the low-frequency limit both DLS-$\mu$r \cite{mason1996rheology} and PT-$\mu$r \cite{Mason:1997yu} have been usefully employed in the past. PT-$\mu$r is technically the less demanding technique, not requiring any laser source or digital correlation board and is also very flexible for biophysical applications, owing to the possibility of employing different sample contrast mechanisms. However, in its practical realization one encounters some challenges. Accurate tracking algorithms require several input parameters, such as a typical value for the particle radius, a score cut-off to discriminate signals that are not due to presence of a particle, an intensity threshold to consider bright pixels as particles, etc. The results of the tracking depends severely on the choice of these parameters that, even for experienced users, may be sometimes more difficult than expected \cite{chenouard2014objective}. Also, the extraction of the tracer MSD from PT trajectories requires the knowledge of the intrinsic particle localization uncertainty, which is usually determined by calibration with particles that are kept fixed in space or that freely diffuse in a Newtonian fluid with similar optical properties \cite{Waigh:2005jk}.

We have shown here that DDM \citep{Cerbino:2008if}, a technique that retains the simplicity and flexibility of PT in terms of experimental setup and applications, can be also used for accurate microrheology experiments. We also show that DDM-$\mu$r outperforms PT with small particles in bright field microscopy. Finally, if an optimization-based algorithm is used instead of the standard fitting-based approach, DDM-$\mu$r does not require any calibration or user input, which limits dramatically the degree of arbitrariness on the determination of the mechanical moduli of the sample. However, particle tracking is expected to be superior to DDM in the presence of unwanted and moving scatterers that, being potentially discarded by an accurate particle tracking, would affect DDM-$\mu$r experiments.

It is likely that these and other DDM features, such as its capability to handle optically dense samples, for which tracking becomes extremely challenging if not impossible, will make DDM-$\mu$r a useful addition to the portfolio of rheo-scientists, both in academic and in industrial research laboratories.

\appendix
\section{Optimization-based determination of MSD}

In this appendix we describe the fitting-free optimization procedure used to extract from the experimental image structure function $D(q,t)$ the best estimate for the tracers' mean square displacement. The main steps of the procedure are the following:
\begin{enumerate}
\item Choice of the interval $[q_1,q_2]$ of wave-vectors over which the optimization is performed. The interval should be a subset of the accessible $q$-range with a fair signal to noise ratio. This condition can be also checked retrospectively at the end of the procedure, when a $q$-resolved estimate of the amplitude $A(q)$ and the noise background $B(q)$ is obtained. 

\item Choice of the initial set of parameters $\left(A_0(q),B_0(q)\right)$. This can be done, for example, by fitting, for each $q \in [q_1,q_2]$, $D(q,t)$ with a linear function near the origin and with a exponential function for large delays (as done for example in Fig.\ref{fig:PEO2rightandwrong}).
\item Calculation of the mean square displacement $MSD(t|q)$ using Eq. \ref{eq:msdddm} for each $q$ in the selected interval.
\item Determination, for each delay time $t$, of the subset $J(t)$ of $q$-values such that $MSD(t|q)<q^{-2}$. This choice ensures that, if $q \in J(t)$, then $D(q,t)$ has not completely lost track of the signal correlation for that value of $q$ and can be thus meaningfully inverted. Let $N(t)$ be the number of elements in $J(t)$.
\item Calculation of the average mean square displacement 
\begin{equation}
MSD(t)=\frac{1}{N(t)}\sum_{q\in J(t)}MSD(t|q).
\label{appMSD}
\end{equation}
\item Calculation of the $t$-dependent dispersion $\sigma_t^2(t)$ as 
\begin{equation}
\sigma_t^2(t)=\frac{1}{N(t)-1}\sum_{q\in J(t)} \log^2  \frac{MSD(t|q)}{MSD(t)}.
\label{appsigma1}
\end{equation}
and of the total dispersion $\sigma^2$ as
\begin{equation}
\sigma^2=\sum_{t}\sigma^2(t).
\label{appsigma2}
\end{equation}
\item Generation of a new set of parameters and repetition of the procedure from step 3 unless a local minimum in $\sigma^2$ is reached (or the prescribed maximum number of iterations is exceeded). 
\item If the procedure converges to a minimum of $\sigma^2$, the optimal set of parameters $\left( A(q),B(q) \right)$ represents the best estimate for the $q$-dependent amplitude and noise baseline, respectively, and the corresponding average mean square displacement (Eq. \ref{appMSD}) is the best estimate for the tracer's MSD.

\end{enumerate}
Many algorithms are available to search the minimum of $\sigma^2$ and to guide the generation of new sets of parameters in step 7. In our implementation, the optimization cycle 3-7 was realized using the MATLAB function \textit{fminsearch}, which is based on the simplex search method of Lagarias \textit{et al.} \cite{fminsearch}. More refined implementation could possibly include suitable weights when computing the averages in the right-hand sides of Eqs. \ref{appMSD}-\ref{appsigma2}, accounting for the different statistical errors affecting each term. Also, an effective weighting scheme could provide an efficient way to reject the contribution of the most noisy wave-vectors making unnecessary the explicit selection of a predetermined optimization interval (step 1).

\begin{acknowledgments}
We thank Stefano Buzzaccaro, Marco Caggioni and Giuliano Zanchetta for stimulating discussions. We thank Gora Conley for careful reading of the manuscript. We acknowledge funding from the Italian Ministry of University and Scientific Research (MIUR) - Project RBFR125H0M, from Regione Lombardia and CARIPLO foundation - Project 2016-0998 and from the Swiss National Science Foundation - Project 200021-157214. 
\end{acknowledgments}

\bibliographystyle{apsrev4-1}

\end{document}